\documentclass[twocolumn,aps,pra,floatfix,superscriptaddress,notitlepage]{revtex4-1}
\usepackage[utf8]{inputenc}
\usepackage[OT1]{fontenc}
\usepackage{amsmath}
\usepackage{amsfonts}
\usepackage{amssymb}
\usepackage{bm}
\usepackage{graphicx}
\usepackage[left=2cm,right=2cm,top=2cm,bottom=2cm]{geometry}
\usepackage{longtable,booktabs}
\allowdisplaybreaks
\newcommand{\bnabla}{\bm{\nabla}}

\usepackage{dcolumn}
\usepackage{siunitx}
\usepackage{multirow}
\usepackage{bigdelim}

\usepackage{color}
\definecolor{BLUE}{rgb}{0.0,0.0,1.0}

\usepackage{soul}
\soulregister\cite7
\soulregister\ref7

\begin{document}

\title{Self-energy corrections to the ionization energies in sodium-like ions: comparison of the \textit{ab initio} QED and model-QED-operator approaches}

\author{P.~Yang}
\affiliation{Innovation Academy for Precision Measurement Science and Technology, Chinese Academy of Sciences, Wuhan 430071, China
\looseness=-1}

\affiliation{University of Chinese Academy of Sciences, Beijing 100049, China
\looseness=-1}

\affiliation{Department of Physics, St.~Petersburg State University, Universitetskaya 7/9, St.~Petersburg 199034, Russia  
\looseness=-1}

\author{A.~V.~Malyshev}
\affiliation{Department of Physics, St.~Petersburg State University, Universitetskaya 7/9, St.~Petersburg 199034, Russia  
\looseness=-1}

\affiliation{Petersburg Nuclear Physics Institute named by B.P. Konstantinov of National Research Center “Kurchatov Institute”, Gatchina, Leningrad region 188300, Russia
\looseness=-2}

\author{E.~A.~Prokhorchuk}
\affiliation{Department of Physics, St.~Petersburg State University, Universitetskaya 7/9, St.~Petersburg 199034, Russia  
\looseness=-1}

\affiliation{School of Physics and Engineering, ITMO University, Kronverkskiy~49, St.~Petersburg 197101, Russia \looseness=-1}

\author{I.~I.~Tupitsyn}
\affiliation{Department of Physics, St.~Petersburg State University, Universitetskaya 7/9, St.~Petersburg 199034, Russia  
\looseness=-1}

\author{V.~M.~Shabaev}
\affiliation{Department of Physics, St.~Petersburg State University, Universitetskaya 7/9, St.~Petersburg 199034, Russia  
\looseness=-1}

\affiliation{Petersburg Nuclear Physics Institute named by B.P. Konstantinov of National Research Center “Kurchatov Institute”, Gatchina, Leningrad region 188300, Russia
\looseness=-2}

\author{D.~P.~Usov}
\affiliation{Department of Physics, St.~Petersburg State University, Universitetskaya 7/9, St.~Petersburg 199034, Russia  
\looseness=-1}

\begin{abstract}
Calculations of the self-energy corrections to ionization energies of the $3s$, $3p_{1/2}$, and $3p_{3/2}$ states in sodium-like  ions with nuclear-charge numbers $Z=30$, 50, 70, and 92 are presented. The calculations are performed using two approaches: the  rigorous bound-state QED formalism and the model-QED-operator method. Within the first method, the first and second orders of the QED perturbation theory formulated in the Furry picture are evaluated. Various screening potentials are included into the initial approximation to partially take into account the electron-electron interaction effects already at the lowest order, thereby accelerating the convergence of perturbation series. Within the second approach, different implementations of the model-QED operator, including its incorporation into the relativistic configuration-interaction calculations, are considered. A detailed comparison of the results obtained by these two independent methods is presented, demonstrating good agreement and thus validating the accuracy and efficiency of the model-QED-operator approach for many-electron systems.

\end{abstract}


\maketitle


\section{Introduction \label{sec:0}}
The self-energy correction constitutes the leading QED contribution in atomic-structure calculations.
For highly charged ions, where the expansion in the electron--nucleus coupling $\alpha Z$ ($\alpha$ is the fine-structure constant and $Z$ is the nuclear-charge number) breaks down, the accurate evaluation of the self-energy contribution is not only crucial but also represents a significant challenge for theory.
Numerical calculations of the self-energy correction to all orders in $\alpha Z$ have a long history and began with pioneering works of Refs.~\cite{Desiderio:1971:1267,Mohr:1974:26,Mohr:1974:52}. Owing to the significance of the self-energy contributions, numerous methods have been proposed in the literature~\cite{Indelicato:1992:172,Persson:1993:125,Quiney:1993:132,Labzowsky:1997:177,Indelicato:1998:165,Jentschura:1999:53}. The most widely used method is the potential-expansion method introduced by Snyderman~\cite{Snyderman:1991:43} and implemented by Blundell and Snyderman~\cite{Snyderman:1991:R1427}, see also Refs.~\cite{Blundell:1992:3762,Cheng:1993:1817,Yerokhin:1999:800}. 
Interest in few-electron systems, such as helium- or lithium-like highly charged ions, has stimulated the development of methods to treat accurately the two-electron self-energy contributions \cite{Persson:1996:204,Yerokhin:1997:361,Yerokhin:1999:3522,Indelicato:2001:052507}. One of the most significant difficulties limiting the numerical accuracy of various \textit{ab initio} methods stems from a slow convergence of partial-wave expansions they involve. However, many approaches have recently been proposed that partially overcome these obstacles \cite{Yerokhin:2005:042502, Artemyev:2007:173004, Sapirstein:2023:042804, Malyshev:2024:062802, Yerokhin:2024:251803, Yerokhin:2025:012802}. 

For highly charged ions, therefore, one can currently calculate both the leading-order and screened self-energy corrections using the rigorous QED approach. However, the $ab~initio$ QED calculations are rather complicated even for one-valence electron and may become extremely difficult for ions  with several valence electrons~\cite{Artemyev:2005:062104,Malyshev:2021:183001,Yerokhin:2022:022815,Malyshev:2025:062811}. Moreover, perturbative treatment of the interelectronic-interaction effects may fail  for neutral systems or those close to them. For these reasons,  simple approximate methods to take into account the radiative corrections are of great importance. Over the past decades, numerous such approaches have been proposed~\cite{Indelicato:1990:5139, Pyykko:2003:1469, Draganic:2003:183001, Flambaum:2005:052115, Thierfelder:2010:062503, Pyykko:2012:371, Tupitsyn:2013:682, Shabaev:2013:012513, Ginges:2016:052509, Skripnikov:2021:201101, Malyshev:2022:012806}. In the present work, we employ the model-QED-operator approach~\cite{Shabaev:2013:012513} realized in the QEDMOD package~\cite{Shabaev:2015:175:2018:69:join_pr}.

The model-QED operator is widely used, but to confirm its applicability and to reinforce its position as a reliable method for incorporating the QED effects,  further comparison between $ab ~initio$ and approximate approaches is necessary, since there are few  such comparisons in the literature \cite{Shabaev:2013:012513, Shabaev:2020:052502}.
The present work is devoted to partially fill this gap. Namely, we study the possibility of applying the model-QED operator to the evaluation of  the self-energy contributions to  ionization energies of ions with the sodium-like electronic configuration: $1s^2 2s^2 2p^6\, v$, where $v$ means the valence electron in the $3s$, $3p_{1/2}$, or $3p_{3/2}$ state. The results are compared with \textit{ab initio} calculations as well as with estimates available in the literature.


Relativistic units ($\hbar=1$ and $c=1$) and the Heaviside charge unit ($e^2=4\pi\alpha$, where $e<0$ is electron charge) are used throughout the paper.

\section{Theoretical approach and computational details \label{sec:1}}
In highly charged ions, the number of electrons $N$ is much smaller than the nuclear-charge number $Z$. Consequently, to zeroth order, both the electron-electron interaction and the coupling to the quantized electromagnetic field can be neglected compared to the electron-nucleus interaction. Therefore, within the Furry picture of QED \cite{Furry:1951:115}, the atomic system is described  under the assumption that the unperturbed one-electron wave functions obey the Dirac equation
\begin{equation}
\label{eq:dirac}
h^{\rm D} \psi \equiv 
\left[ \bm{\alpha} \cdot \bm{p} + \beta m + V \right] \psi = 
\varepsilon\psi \, ,
\end{equation}
where $\bm{p}=-i\bnabla$ is the momentum operator and $\bm{\alpha}$ and $\beta$ are the Dirac matrices.
In the simplest case, $V(r)$ in Eq.~(\ref{eq:dirac}) represents the nuclear potential $V_{\rm nucl}(r)$. In this work, we choose $V(r)$ to be an effective potential, namely, the sum of the nuclear potential and a screening potential:
\begin{equation}
\label{eq:V(r)}
V(r)\rightarrow V_{{\rm eff}}(r)=V_{{\rm nucl}}(r)+V_{{\rm scr}}(r) \, .
\end{equation}
Screening potentials serve to partially model the interelectronic-interaction effects already in the zeroth-order approximation. This modification leads to a rearrangement of the perturbation series, since the counterterm $\delta V=-V_{\rm scr}$ must be considered perturbatively in this case, thus improving its convergence.
In the present work we employ a number of screening
potentials. 
Most of them can be expressed using the charge density of the core electrons
\begin{equation}
\label{eq:rho_core}
\rho_{\rm core}(r)=\sum_c(2j_c+1)\left[G_c^2(r)+F_c^2(r)\right] \,,
\end{equation}
where the index $c$ runs over all closed-shell states ($1s$, $2s$, $2p_{1/2}$, and $2p_{3/2}$), $j_c$ is the total angular momentum of the Dirac wave function $\psi_c$, and $G_c$ and $F_c$ are its large and small components, normalized such that $\int\!dr\left[G^2_c(r)+F^2_c(r)\right]=1$.
Then, the screening potential reads as
\begin{equation}
\label{eq:V_scr}
V_{{\rm scr}}(r)=\alpha{\rm \int_0^{\infty}}\!dr'\frac{\rho_{\rm core}(r')}{\max\{r,r'\}}-x_{a}\frac{\alpha}{r}\left(\frac{81}{32\pi^2}r\rho_{\rm core}(r)\right)^{1/3} \,.
\end{equation}
In our calculations, we use the screening potentials (\ref{eq:V_scr}) with $x_a =0$, $1/3$, $2/3$, and $1$ and refer to them as ``x0'', ``x1'', ``x2'', and ``x3'', respectively. In particular, $x_{a}=0$ is the core-Hartree potential, while $x_{a}=2/3$ and $x_{a}=1$ are the Kohn-Sham \cite{pot:KS} and Dirac-Slater \cite{Slater:1951:385} potentials for the $1s^2 2s^2 2p^6$ electronic configuration. In addition, to enable a direct comparison with the results obtained by Sapirstein and Cheng in Ref.~\cite{2015}, we also adopt the screening potential as in their work. Specifically, we construct the Kohn-Sham potential (i.e., $x_a=2/3$) by replacing $\rho_{\rm core}(r)$ in Eq.~(\ref{eq:V_scr})  with the total charge density
\begin{equation}
\label{eq:rho_sc}
\rho_{\rm tot}(r) = \rho_{\rm core}(r) + \left[G_{3s}^2(r)+F_{3s}^2(r)\right] \,.
\end{equation}
In what follows, we refer to this potential as ``SC''. All the potentials are constructed by self-consistently solving the Dirac equation (\ref{eq:dirac}), without introducing the Latter correction \cite{Latter:1955:510}. Therefore, for large $r$ the $x_a$-potentials behave as $\alpha (N-1)/r$, whereas the SC potential behaves as $\alpha N/r$.

\begin{figure}[ht]
  \centering
  \includegraphics[height=0.12\textheight]{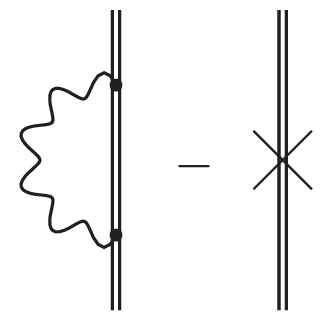}
  \caption{First-order self-energy diagram. The double line denotes the bound-electron propagator, the wavy line corresponds to the photon propagator, and the cross indicates the mass counterterm.}
  \label{fig:se_1el}
\end{figure}

The one-electron self-energy Feynman diagram, together with the associated mass counterterm, is shown in Fig.~\ref{fig:se_1el}.
The corresponding energy shift  for a bound electron in the state $\psi_a$ is given by the real part of the  expression
\begin{equation}
\begin{aligned}
\Delta E_a^{(1)}=&2i\alpha{\rm \int_{-\infty}^{\infty}}\!d\omega{\rm \int}\!d\bm{r}_1{\rm \int}\!d\bm{r}_2\,\psi_a^{\dagger}(\bm{r}_1)\alpha_{\mu}\\
&\times G(\varepsilon_a-\omega,\bm{r}_1,\bm{r}_2)\alpha_{\nu}\psi_a(\bm{r}_2)D^{\mu\nu}(\omega,\bm{r}_{12})\\
&-\delta m{\rm \int}\!d\bm{r}\,\psi_a^{\dagger}(\bm{r})\beta\psi_a(\bm{r}) \equiv\langle \psi_a | \Sigma_R(\varepsilon_a)|\psi_a\rangle \, ,
\end{aligned}
\end{equation}
where $\varepsilon_a$ is the Dirac energy of the state $\psi_a$, $\alpha^{\mu}=(1,\bm{\alpha})$, $\bm{r}_{12}=\bm{r}_1-\bm{r}_2$, $D^{\mu\nu}(\omega,\bm{r}_{12})$ is the photon propagator, $G(\omega,\bm{r}_1,\bm{r}_2)$ is the bound-electron Green's function,  $\delta m$ is the mass counterterm, and we have defined the renormalized self-energy operator $\Sigma_R(\varepsilon)$. 

To handle the ultraviolet divergences encountering in the self-energy correction, we employ the potential-expansion approach, decomposing the total contribution into three components: 
\begin{equation}
\label{eq:se_1st}
\Delta E_a^{(1)}=\Delta E_{\rm zero}+\Delta E_{\rm one}+\Delta E_{\rm many} \,.
\end{equation}
The zero- and one-potential terms are evaluated in momentum space after applying the renormalization procedure, while the many-potential term is computed in coordinate space using the partial-wave expansions for $G$ and $D^{\mu\nu}$. The specific details can be found, e.g., in Ref.~\cite{Yerokhin:1999:800}. To improve the convergence of the partial-wave expansion in the calculations of the one-electron self-energy corrections, we use the method from Ref.~\cite{Sapirstein:2023:042804}.

\begin{figure}[ht]
  \centering
  \includegraphics[height=0.12\textheight]{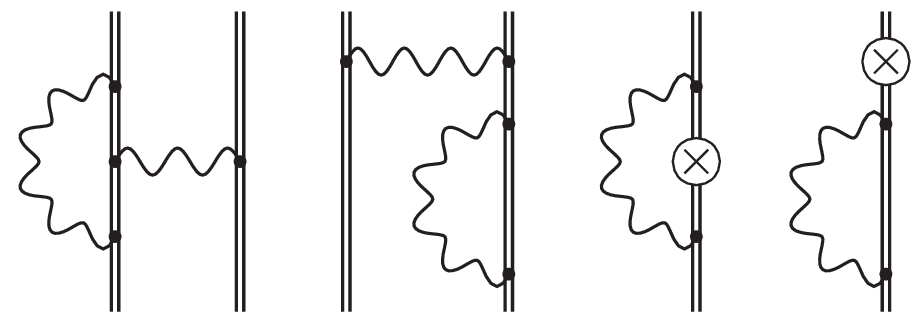}
  \caption{Second-order self-energy diagram. The circle with a cross inside represents the local screening potential counterterm $\delta V=-V_{\rm scr}$. The other notations are the same as in Fig.~\ref{fig:se_1el}. The mass counterterm diagrams are not shown. }
  \label{fig:se_2el}
\end{figure}

The treatment of ionization energies for sodium-like ions requires evaluating the first-order self-energy contribution (\ref{eq:se_1st}) for the valence electrons. Within the rigorous QED framework, the corresponding radiative corrections can be further refined by taking into account the second-order Feynman diagrams shown in Fig.~\ref{fig:se_2el}. In the present work, we perform such calculations. All the necessary formulas, can be found, e.g., in Refs.~\cite{Yerokhin:1999:3522,Malyshev:2017:022512}. The convergence of partial-wave expansions is improved in this case using the approach described in Ref. \cite{Malyshev:2024:062802}.

The model-QED-operator approach is developed within the framework of the two-time Green's function (TTGF) method~\cite{TTGF}, which also represents the basis for our \textit{ab initio} QED calculations. The TTGF method provides a systematic derivation of  effective Hamiltonians that incorporate QED corrections through its action in a suitably defined active space~\cite{Shabaev:1993:4703, TTGF}.
Within this approach~\cite{Shabaev:2013:012513}, the self-energy (SE) operator is represented as a sum of semilocal and nonlocal parts,
\begin{equation}
\label{eq:se_model_1}
h^{\rm SE} = h^{\rm SE}_{\rm sl} + h^{\rm SE}_{\rm nl} \, .
\end{equation}
The semilocal operator $h^{\rm SE}_{\rm sl}$ acts differently on wave functions of different angular symmetry, and for a given relativistic angular quantum number $\kappa=(-1)^{j+l+1/2}(j+1/2)$ it takes the form
\begin{equation}
\label{eq:se_model_2}
h^{\rm SE}_{\rm sl,\kappa}=A_{\kappa}\exp(-r/\lambdabar_{\rm C}) \,.
\end{equation}
Here, $\lambdabar_{\rm C}=1/m$ is the Compton wavelength and the constant $A_{\kappa}$ is determined by requiring that the matrix element of $h^{\rm SE}_{\rm sl,\kappa}$, evaluated with the hydrogenlike wave function of the lowest energy state for the corresponding $\kappa$, reproduces the exact self-energy correction. ``Hydrogenlike'' refers here to the solutions of the Dirac equation (\ref{eq:dirac}) with the pure nuclear potential $V=V_{\rm nucl}$. The nonlocal operator $h^{\rm SE}_{\rm nl}$ is represented in a separable form,
\begin{equation}
\label{eq:se_model_3}
h^{\rm SE}_{\rm nl} = \sum_{i,k=1}^n 
| \phi_i \rangle B_{ik} \langle \phi_k | \, ,
\end{equation}
where the specific choice of the projector functions $\{\phi_i\}_{i=1}^n$ is discussed in details in Refs.~\cite{Shabaev:2013:012513}.
The constants $B_{ik}$ are determined by the requirement that the diagonal and off-diagonal matrix elements of $h^{\rm SE}$, evaluated in the basis of hydrogenlike wave functions $\{\psi_i^{\rm H}\}_{i=1}^n$, reproduce the exact first-order self-energy contributions,
\begin{equation}
\label{eq:se_model_4}
\langle\psi_i^{\rm H} | h^{\rm SE} | \psi_k^{\rm H} \rangle = \frac{1}{2} \langle \psi_i^{\rm {H}} | \left[\Sigma_{{R}}(\varepsilon_i) + \Sigma_{{R}}(\varepsilon_k)\right] | \psi_k^{\rm {H}}\rangle\, .
\end{equation} 

In the present work, we employ the model-QED operator in several different ways to approximately evaluate the self-energy contribution to the ionization energies of sodium-like ions.  The simplest approach, which we refer to as ``QEDMOD(av)'', consists of calculating the matrix element of $h^{\rm SE}$ with the valence-electron wave functions obtained from the Dirac equation~(\ref{eq:dirac}) which includes the various screening potentials.
More sophisticated approaches involve incorporating the model-QED operator into the relativistic correlation
calculations based on the Dirac-Coulomb-Breit (DCB) Hamiltonian.

The DCB Hamiltonian can be expressed as
\begin{align}
\label{eq:se_model_6}
H^{\rm DCB} &= \Lambda^{(+)}\left[
{H_0 + V_{\rm int}}
\right]\Lambda^{(+)} \, , \\
{H_0} &{= \sum_i \left[ \bm{\alpha}_i \cdot \bm{p}_i + \beta_i m + V_{{\rm nucl,}i} \right]} \, , \\
{V_{\rm int}} &{= \sum_{i<j} \left[ V^{\rm C}_{ij} + V^{\rm B}_{ij} \right]  \,,}
\end{align} 
where the sums run over all the atomic electrons
and the Coulomb and Breit parts of the interelectronic-interaction operator read as
\begin{align}
\label{eq:V_C}
V^{\rm C}_{ij} &= \frac{\alpha}{r_{ij}} \, , \\ 
\label{eq:V_B}
V^{\rm B}_{ij} &= -\frac{\alpha}{2} \left[ \frac{\bm{\alpha}_i \cdot \bm{\alpha}_j}{r_{ij}}
+ \frac{(\bm{\alpha}_i \cdot \bm{r}_{ij})(\bm{\alpha}_j \cdot \bm{r}_{ij})}{r_{ij}^3}  \right] \, .
\end{align}
The operator $\Lambda^{(+)}$ in Eq.~(\ref{eq:se_model_6}) is defined as the projector onto the space of many-electron states constructed from the positive-energy eigenfunctions of some one-particle Hamiltonian $h^{\Lambda}$. In principle, the operator $h^{\Lambda}$ can be chosen in many different ways. For example, one can use the Dirac Hamiltonian, which incorporates either the pure Coulomb or an effective potential or the nonlocal DF operator $h^{\rm DF}$. For the specific choices, used in the present paper, see the discussion below.

To solve the DCB equation
\begin{eqnarray}
H^{\rm DCB} \Psi = E \Psi \,,
\end{eqnarray}
we use the configuration-interaction (CI)  method in the basis of one-electron Dirac-Sturm orbitals~\cite{Bratzev:1977:173,Tupitsyn:2003:022511,Tupitsyn:2018:022517}.
Within this method, the many-electron wave function $\Psi(JM)$, where $J$ denotes the total angular momentum and $M$ stands for its projection, is expressed as an expansion over a large set of configuration-state functions $\{\Phi_I\}$ with the same angular quantum numbers:
\begin{equation}
\label{eq:se_model_8}
{\Psi(JM) = \sum_I C_I \Phi_I(JM) \,.}
\end{equation} 
The functions $\Phi_I$ are given by linear combinations of Slater determinants constructed from the eigenfunctions $\{\varphi^\Lambda_i\}$ of the operator $h^\Lambda$. Finally, the functions $\varphi^\Lambda_i$ are determined within the finite basis consisting of the solutions of the Dirac equation (\ref{eq:dirac}) for occupied orbitals and the solutions of the Dirac-Sturm equation,
\begin{equation}
\label{eq:se_model_9}
\left[ h^{\rm D} - \varepsilon_{0}\right] {\psi}^{\rm DS}_j = \lambda_j W(r){\psi}^{\rm DS}_j \, ,
\end{equation} 
for vacant orbitals.
In Eq. (\ref{eq:se_model_9}), $\varepsilon_0$  denotes a reference one-electron energy and $W(r)$ is a weight function of constant sign given by 
\begin{equation}
\label{eq:se_model_10}
W(r) = \frac{1-\exp[-(\mu r)^2]}{(\mu r)^2} \, .
\end{equation} 
The specific values of $\varepsilon_0$ and  $\mu$ are chosen to minimize the basis-set size required to achieve a convergence of the results.
Within the CI method, the ionization energy of the valence electron in the $v$ state can be obtained as the difference between the total binding energies of the configurations $1s^2 2s^2 2p^6 v$ and $1s^2 2s^2 2p^6$.

The second approach employed in this study, which we refer to as ``CI-QEDMOD'', involves adding the term $\sum_i h_i^{\rm SE}$ to the DCB Hamiltonian. Within this approach, we adopt the one-electron Dirac Hamiltonian $h^{\rm D}$ as the operator $h^\Lambda$. The self-energy contribution is evaluated as the difference between the ionization energies obtained with and without including the model-QED operator in Eq.~(\ref{eq:se_model_6}).

As shown in Ref.~\cite{Shabaev:2020:052502}, the QED contribution can be very sensitive to the interelectronic interaction. Consequently, a more comprehensive inclusion of the model-QED operator into the computational framework can be necessary.
To this end, we also study an approach referred to as ``CI-QEDMOD(scf)'', in which we additionally include the model-QED operator into the operator $h^\Lambda$, i.e., adopt $h^\Lambda=h^{\rm D}+h^{\rm SE}$. 
This inclusion implies a self-consistent modification of the projector operators $\Lambda^{(+)}$ in Eq.~(\ref{eq:se_model_6}).
Compared to ``CI-QEDMOD'', the ``CI-QEDMOD(scf)'' approach also takes into account approximately the  single-particle excitation into the negative-energy continuum.

\section{Numerical results and discussions \label{sec:2}}

In this section, we present the results of our calculations of the self-energy contributions to the ionization energies of sodium-like ions with $Z=30$, 50, 70, and 92. Specifically, we consider the corrections to the binding energy of the valence electron $v$ in the configurations $1s^2 2s^2 2p^6 \,v$ for $v=3s$, $3p_{1/2}$, and $3p_{3/2}$. The calculations are performed both within the rigorous QED approach and using the model-QED operator. We compare the results obtained with these independent methods and study how they depend on the choice of the screening potential that defines the initial approximation. For this aim, we perform the calculations using the SC potential, previously employed in Ref.~\cite{2015}, and the set of the $x_a$-potentials x0, x1, x2, and x3, described in the previous section.

The results for the $3s$, $3p_{1/2}$, and $3p_{3/2}$ states are summarized in
 Tables~\ref{tab:se_3s1_2}, \ref{tab:se_3p1_2}, and \ref{tab:se_3p3_2}, respectively. All results are presented in atomic units. For each state and potential, the rows labeled ``SE'' display the results of our \textit{ab initio} calculations of the first-order one-electron self-energy corrections shown in Fig.~\ref{fig:se_1el}. For the SC potential, we compare our results with those from Ref.~\cite{2015} and find good agreement. The rows labeled ``QEDMOD(av)'' show the expectation values of the model-QED operator $h^{\rm SE}$, computed with the valence-electron wave functions obtained for the corresponding screening potentials. From Tables \ref{tab:se_3s1_2}, \ref{tab:se_3p1_2}, and \ref{tab:se_3p3_2}, we observe reasonable agreement between the results of the rigorous and approximate calculations. We note that the one-electron correction for the $3p_{1/2}$ state is small and crosses zero in the range between $Z=30$ and $Z=50$. Nevertheless, the model-QED operator yields the correct order of magnitude even in this region. We also stress the strong dependence of the obtained results on the choice of the initial approximation.

In the rows labeled ``SE+ScrSE'', the second-order self-energy corrections from Fig.~\ref{fig:se_2el}, obtained within the rigorous QED approach, are added to the first-order results. The numbers in parentheses represent the purely numerical uncertainties arising from the truncation and extrapolation of the partial-wave expansions in the corresponding contributions \cite{Malyshev:2024:062802}. One can see that the two-electron corrections significantly affect the total values of self-energy contributions, especially for small $Z$ and the $3p$ states. This demonstrates the importance of accounting for correlation effects in the treatment of the QED contributions. For the SC potential, we compare our results with those reported in Ref.~\cite{2015}. In this case, however, we find a discrepancy whose origin remains unclear for us. Meanwhile, the correctness of our \textit{ab initio} calculations is supported by the independent ``CI-QEDMOD'' method. This approach incorporates the model-QED operator into the configuration-interaction evaluation of the correlation effects and leads to results that are in good agreement with the ones obtained by the rigorous approach. This is especially evident from the data for $Z=92$, where perturbation theory in $1/Z$ is expected to work well. Indeed, for high $Z$, the addition of the second-order corrections considerably reduces the scatter of the results obtained with the different screening potentials, compared to that of the first-order results. As $Z$ decreases, this scatter increases, while the ``CI-QEDMOD'' values remain close to each other. This is because the configuration-interaction method treats the correlation effects to all orders in $1/Z$, and is thus able to provide reasonable estimates for the self-energy corrections for all considered $3s$, $3p_{1/2}$, and $3p_{3/2}$ states, even when the nuclear-charge number $Z$ and the number of electrons $N$ become comparable. For the configuration-interaction calculations, we do not indicate any numerical uncertainty, as convergence of the results with respect to the number of partial waves and basis size has been achieved.

Finally, comparing the results in the rows labeled ``CI-QEDMOD'' and ``CI-QEDMOD(scf)'' shows that for sodium-like ions considered here, both approaches yield the similar results. It is worth noting, however, that this outcome could not have been easily predicted in advance \cite{Shabaev:2020:052502}.

\begin{table*}[t]
\renewcommand{\arraystretch}{1.2}
\centering
\setlength{\tabcolsep}{0.5cm}
\caption{Self-energy correction for the ionization energy of the $3s$ state in Na-like ions (in a.u.).}
\label{tab:se_3s1_2}
\begin{tabular}{@{\,\,}c@{\quad}l@{\qquad}S[table-format=1.6,group-separator=]S[table-format=1.5,group-separator=]S[table-format=1.4,group-separator=]S[table-format=1.4,group-separator=]@{\,\,}}
\hline
\hline
Potential\rule{0pt}{2.6ex}~ & 
                  ~~~Method & 
                  \multicolumn{1}{l}{\!\!\!\!\!\!\!$Z=30$} & 
                  \multicolumn{1}{l}{\hphantom{-}$Z=50$} & 
                  \multicolumn{1}{l}{\hphantom{-}$Z=70$} & 
                  \multicolumn{1}{l}{\hphantom{-}$Z=92$\!\!\!\!\!\!} \\
\hline
 & SE\rule{0pt}{2.6ex}   & 0.006251 & 0.04794 & 0.1844 & 0.6101 \\
 & SE~\cite{2015} & 0.00625 & 0.0479 & 0.1844 & 0.6101 \\
 & QEDMOD(av) & 0.006252 & 0.04794 & 0.1844 & 0.6095 \\
SC & SE+ScrSE & 0.005711 & 0.04539 & 0.1765 & 0.5860 \\
 & SE+ScrSE~\cite{2015} & 0.00494 & 0.0444 & 0.1753 & 0.5844 \\
 & CI-QEDMOD & 0.005786 & 0.04560 & 0.1771 & 0.5866 \\
 & CI-QEDMOD(scf) & 0.005785 & 0.04565 & 0.1774 & 0.5903 \\
\hline
 & SE\rule{0pt}{2.6ex} & 0.006072 & 0.04709 & 0.1819 & 0.6029 \\
 & QEDMOD(av) & 0.006065 & 0.04704 & 0.1817 & 0.6013 \\
x0 & SE+ScrSE & 0.005736 & 0.04546 & 0.1767 & 0.5865 \\
 & CI-QEDMOD & 0.005785 & 0.04560 & 0.1770 & 0.5865 \\
 & CI-QEDMOD(scf) & 0.005785 & 0.04565 & 0.1774 & 0.5903 \\
\hline
 & SE\rule{0pt}{2.6ex} & 0.006284 & 0.04800 & 0.1844 & 0.6095 \\
 & QEDMOD(av) & 0.006282 & 0.04797 & 0.1843 & 0.6084 \\
x1 & SE+ScrSE & 0.005717 & 0.04541 & 0.1765 & 0.5862 \\
 & CI-QEDMOD & 0.005785 & 0.04560 & 0.1771 & 0.5865 \\
 & CI-QEDMOD(scf) & 0.005785 & 0.04565 & 0.1774 & 0.5903 \\
\hline
 & SE\rule{0pt}{2.6ex} & 0.006500 & 0.04891 & 0.1869 & 0.6162 \\
 & QEDMOD(av) & 0.006501 & 0.04891 & 0.1869 & 0.6156 \\
x2 & SE+ScrSE & 0.005690 & 0.04534 & 0.1764 & 0.5858 \\
 & CI-QEDMOD & 0.005785 & 0.04560 & 0.1771 & 0.5866 \\
 & CI-QEDMOD(scf) & 0.005785 & 0.04565 & 0.1774 & 0.5903 \\
\hline
 & SE\rule{0pt}{2.6ex} & 0.006721 & 0.04984 & 0.1895 & 0.6230 \\
 & QEDMOD(av) & 0.006724 & 0.04985 & 0.1895 & 0.6229 \\
x3 & SE+ScrSE & 0.005652 & 0.04525 & 0.1762 & 0.5853 \\
 & CI-QEDMOD & 0.005786 & 0.04560 & 0.1771 & 0.5866 \\
 & CI-QEDMOD(scf) & 0.005785 & 0.04565 & 0.1774 & 0.5903 \\
\hline
\hline
\end{tabular}
\end{table*}

\begin{table*}[t]
\renewcommand{\arraystretch}{1.2}
\centering
\setlength{\tabcolsep}{0.5cm}
\caption{Self-energy correction for the ionization energy of the $3p_{1/2}$ state in Na-like ions (in a.u.).}
\label{tab:se_3p1_2}
\begin{tabular}{l@{\quad}l@{\quad}S[table-format=2.6,group-separator=]S[table-format=2.5,group-separator=]S[table-format=2.4,group-separator=]S[table-format=2.4,group-separator=]}
\hline
\hline
Potential\rule{0pt}{2.6ex} & 
                  Method & 
                  \multicolumn{1}{l}{\hphantom{-}\!\!\!\!\!\!\!\!$Z=30$} & 
                  \multicolumn{1}{l}{\hphantom{-}$Z=50$} & 
                  \multicolumn{1}{l}{\hphantom{-}$Z=70$} & 
                  \multicolumn{1}{l}{\hphantom{-}$Z=92$} \\
\hline
 & SE\rule{0pt}{2.6ex} & -0.0001498 & 0.000365 & 0.012018 & 0.09940 \\
 & SE~\cite{2015} & -0.00015 & 0.0004 & 0.0120 & 0.0994 \\
 & QEDMOD(av) & -0.0002194 & -0.000101 & 0.010744 & 0.09727 \\
SC & SE+ScrSE & -0.0006233(4) & -0.002009(2) & 0.004250(2) & 0.07382 \\
 & SE+ScrSE~\cite{2015} & -0.00094 & -0.0022 & 0.0043 & 0.0736 \\
 & CI-QEDMOD & -0.0006465 & -0.002375 & 0.003453 & 0.07441 \\
 & CI-QEDMOD(scf) & -0.0006397 & -0.002232 & 0.004210 & 0.07742 \\
\hline
 & SE\rule{0pt}{2.6ex} & -0.0001450 & 0.000326 & 0.011588 & 0.09674 \\
 & QEDMOD(av) & -0.0002385 & -0.000284 & 0.009971 & 0.09422 \\
x0 & SE+ScrSE & -0.0006174(4) & -0.001977(2) & 0.004363(2) & 0.07425 \\
 & CI-QEDMOD & -0.0006466 & -0.002375 & 0.003454 & 0.07444 \\
 & CI-QEDMOD(scf) & -0.0006396 & -0.002232 & 0.004210 & 0.07741 \\
\hline
 & SE\rule{0pt}{2.6ex} & -0.0001497 & 0.000356 & 0.011917 & 0.09870 \\
 & QEDMOD(av) & -0.0002342 & -0.000188 & 0.010466 & 0.09637 \\
x1 & SE+ScrSE & -0.0006294(4) & -0.002015(2) & 0.004261(2) & 0.07393 \\
 & CI-QEDMOD & -0.0006466 & -0.002375 & 0.003453 & 0.07442 \\
 & CI-QEDMOD(scf) & -0.0006397 & -0.002232 & 0.004210 & 0.07742 \\
\hline
 & SE\rule{0pt}{2.6ex} & -0.0001546 & 0.000387 & 0.012252 & 0.10069 \\
 & QEDMOD(av) & -0.0002289 & -0.000088 & 0.010969 & 0.09854 \\
x2 & SE+ScrSE & -0.0006413(4) & -0.002054(2) & 0.004148(2) & 0.07357 \\
 & CI-QEDMOD & -0.0006465 & -0.002375 & 0.003453 & 0.07441 \\
 & CI-QEDMOD(scf) & -0.0006397 & -0.002232 & 0.004210 & 0.07742 \\
\hline
 & SE\rule{0pt}{2.6ex} & -0.0001594 & 0.000420 & 0.012595 & 0.10271 \\
 & QEDMOD(av) & -0.0002225 & 0.000015 & 0.011481 & 0.10075 \\
x3 & SE+ScrSE & -0.0006532(4) & -0.002096(2) & 0.004026(2) & 0.07316 \\
 & CI-QEDMOD & -0.0006464 & -0.002375 & 0.003452 & 0.07440 \\
 & CI-QEDMOD(scf) & -0.0006397 & -0.002232 & 0.004210 & 0.07742 \\
\hline
\hline
\end{tabular}
\end{table*}

\begin{table*}[t]
\renewcommand{\arraystretch}{1.2}
\centering
\setlength{\tabcolsep}{0.5cm}
\caption{Self-energy correction for the ionization energy of the $3p_{3/2}$ state in Na-like ions (in a.u.).}
\label{tab:se_3p3_2}
\begin{tabular}{l@{\quad}l@{\quad}S[table-format=2.6,group-separator=]S[table-format=2.5,group-separator=]S[table-format=2.4,group-separator=]S[table-format=2.4,group-separator=]}
\hline
\hline
Potential\rule{0pt}{2.6ex} & 
                  Method & 
                  \multicolumn{1}{l}{\!\!\!\!\!\!\!$Z=30$} & 
                  \multicolumn{1}{l}{\hphantom{-}$Z=50$} & 
                  \multicolumn{1}{l}{\hphantom{-}$Z=70$} & 
                  \multicolumn{1}{l}{\hphantom{-}$Z=92$\!\!\!\!\!\!} \\
\hline
 & SE\rule{0pt}{2.6ex} & 0.0002675 & 0.003975 & 0.02197 & 0.08794 \\
 & SE~\cite{2015} & 0.00027 & 0.0040 & 0.0220 & 0.0879 \\
 & QEDMOD(av) & 0.0003039 & 0.004140 & 0.02239 & 0.08842 \\
SC & SE+ScrSE & -0.0002034(4) & 0.001808(2) & 0.01587 & 0.07246 \\
 & SE+ScrSE~\cite{2015} & -0.00067 & 0.0014 & 0.0154 & 0.0718 \\
 & CI-QEDMOD & -0.0001078 & 0.002113 & 0.01663 & 0.07377 \\
 & CI-QEDMOD(scf) & -0.0001104 & 0.002107 & 0.01661 & 0.07375 \\
\hline
 & SE\rule{0pt}{2.6ex} & 0.0002526 & 0.003848 & 0.02147 & 0.08644 \\
 & QEDMOD(av) & 0.0002850 & 0.004000 & 0.02188 & 0.08693 \\
x0 & SE+ScrSE & -0.0001965(4) & 0.001836(2) & 0.01594 & 0.07262 \\
 & CI-QEDMOD & -0.0001078 & 0.002113 & 0.01663 & 0.07377 \\
 & CI-QEDMOD(scf) & -0.0001104 & 0.002107 & 0.01661 & 0.07375 \\
\hline
 & SE\rule{0pt}{2.6ex} & 0.0002669 & 0.003962 & 0.02191 & 0.08774 \\
 & QEDMOD(av) & 0.0003012 & 0.004119 & 0.02232 & 0.08823 \\
x1 & SE+ScrSE & -0.0002085(4) & 0.001803(2) & 0.01587 & 0.07247 \\
 & CI-QEDMOD & -0.0001078 & 0.002113 & 0.01663 & 0.07377 \\
 & CI-QEDMOD(scf) & -0.0001104 & 0.002107 & 0.01661 & 0.07375 \\
\hline
 & SE\rule{0pt}{2.6ex} & 0.0002819 & 0.004079 & 0.02235 & 0.08906 \\
 & QEDMOD(av) & 0.0003181 & 0.004241 & 0.02276 & 0.08954 \\
x2 & SE+ScrSE & -0.0002217(4) & 0.001766(2) & 0.01579 & 0.07230 \\
 & CI-QEDMOD & -0.0001077 & 0.002113 & 0.01663 & 0.07377 \\
 & CI-QEDMOD(scf) & -0.0001104 & 0.002107 & 0.01661 & 0.07375 \\
\hline
 & SE\rule{0pt}{2.6ex} & 0.0002974 & 0.004198 & 0.02280 & 0.09040 \\
 & QEDMOD(av) & 0.0003359 & 0.004366 & 0.02321 & 0.09087 \\
x3 & SE+ScrSE & -0.0002362(4) & 0.001724(2) & 0.01570 & 0.07210 \\
 & CI-QEDMOD & -0.0001077 & 0.002113 & 0.01663 & 0.07377 \\
 & CI-QEDMOD(scf) & -0.0001104 & 0.002107 & 0.01661 & 0.07375 \\

\hline
\end{tabular}
\end{table*}

\section{Summary \label{sec:3}}
The self-energy contributions to the ionization energies of sodium-like ions with $Z=30$, 50, 70, and 92 are evaluated. The calculations are performed using two independent approaches: a rigorous QED treatment up to the second order of perturbation theory and the approximate methods based on the model-QED operator. The dependence of the results on the choice of the initial approximation is studied by varying the local screening potential incorporated into the unperturbed Hamiltonian. While a comparison of the results from the two methods shows good agreement, a  dependence on the zeroth-order approximation for the QED perturbation theory is revealed when $Z$ decreases. This dependence stems from the significant correlation effects, which can hardly be described properly by perturbation theory. It is demonstrated that combining the model-QED-operator approach with the configuration-interaction method yields results that are almost insensitive to this choice. The obtained results confirm that the corresponding computational scheme provides a reliable approach for evaluating the self-energy effects in many-electron systems.


\section*{Acknowledgments}

This work was funded by China Scholarship Council (CSC, Grant No. 202404910570).
The work of A.~V.~M., E.~A.~P., and D.~P.~U. was supported by the Foundation for the Advancement of
Theoretical Physics and Mathematics BASIS (Project No. 24-1-2-74-1).

\end{document}